\documentclass[a4paper,pre,twocolumn]{revtex4-1}
\usepackage[dvips]{graphicx}
\usepackage{amsmath}
\usepackage{psfrag,color}
\usepackage{amssymb}
\usepackage{natbib}
\usepackage{ulem}
\newcommand{\dL}{\mbox{$\Delta L$}}

\newcommand{\rhos}{\rho_s}
\newcommand{\tast}{t_\ast}

\newcommand{\Lf}{\mbox{$\Delta L_{\mathrm{fold}}$}}
\newcommand{\wf}{w_{\mathrm{fold}}}
\newcommand{\Wf}{W_{\mathrm{fold}}}
\newcommand{\muf}{\mu_{\mathrm{fold}}}
\newcommand{\tauf}{\tau_{\mathrm{fold}}}
\newcommand{\tb}{t_{\mathrm{snap}}}
\newcommand{\tbot}{{\cal T}_{\mathrm{b}}}
\newcommand{\Dmu}{\Delta \mu}

\newcommand{\tnd}{\mathcal{T}}

\newcommand{\tr}{t_{\mathrm{raw}}}
\newcommand{\Wt}{W}
\newcommand{\taut}{\tau}

\newcommand \beq{\begin{equation}}
\newcommand \eeq{\end{equation}}
\newcommand \beqn{\begin{equation*}}
\newcommand \eeqn{\end{equation*}}

\newcommand{\id}{\mathrm{d}} 
\renewcommand{\d}[2]{\frac{\id #1}{\id #2}} 
\newcommand{\dd}[2]{\frac{\id^2 #1}{\id #2^2}} 

\newcommand{\pd}[2]{\frac{\partial #1}{\partial #2}} 
\newcommand{\pdd}[2]{\frac{\partial^2 #1}{\partial #2^2}} 
\newcommand{\pdf}[2]{\frac{\partial^4 #1}{\partial #2^4}} 

\begin{document}
\title{Critical slowing down in purely elastic `snap-through' instabilities}

\author{Michael Gomez, Derek E.~Moulton and Dominic Vella}
\email[]{dominic.vella@maths.ox.ac.uk}
\affiliation{ Mathematical Institute, University of Oxford, Woodstock Rd, Oxford, OX2 6GG, UK}


\maketitle


\textbf{
Many elastic structures have two possible equilibrium states \cite{Bazant}: from  umbrellas that become inverted in a sudden gust of wind, to nano-electromechanical switches \cite{Loh2012,Xu2014}, origami patterns \cite{Silverberg2015,Dudte2016} and the hopper popper, which jumps after being turned inside-out \cite{pandey_dynamics_2014}. These systems typically transition from one state to the other via a rapid `snap-through'. Snap-through allows plants to   gradually store elastic energy, before releasing it suddenly to generate rapid motions \cite{Skotheim2005,Forterre2013}, as in the Venus flytrap \cite{Forterre2005}. Similarly, the beak of the hummingbird snaps through to catch insects mid-flight \cite{Smith2011}, while technological applications are increasingly exploiting snap-through instabilities \cite{Goncalves2003,Daynes2010,Hung1999}. In all of these scenarios, it is the ability to repeatedly generate fast motions that gives snap-through its utility. However, estimates of the speed of snap-through suggest that it should occur more quickly than is usually observed. Here, we study the dynamics of snap-through in detail, showing that, even without dissipation, the dynamics slow down close to the snap-through transition. This is reminiscent of the slowing down observed in critical phenomena, and provides a handheld demonstration of such phenomena, as well as a new tool for tuning dynamic responses in applications of elastic bistability. 
}

Snap-through occurs when a system is in an equilibrium state that either ceases to exist or becomes unstable as a control parameter varies: the system must jump to another equilibrium state. For example, in the Venus flytrap it is believed that the natural curvature of the leaf changes slightly making the `open' equilibrium disappear leaving only the `closed' equilibrium \cite{Forterre2005}.

\begin{figure*}
\centering
\hspace{30pt}
\includegraphics[width =  2\columnwidth]{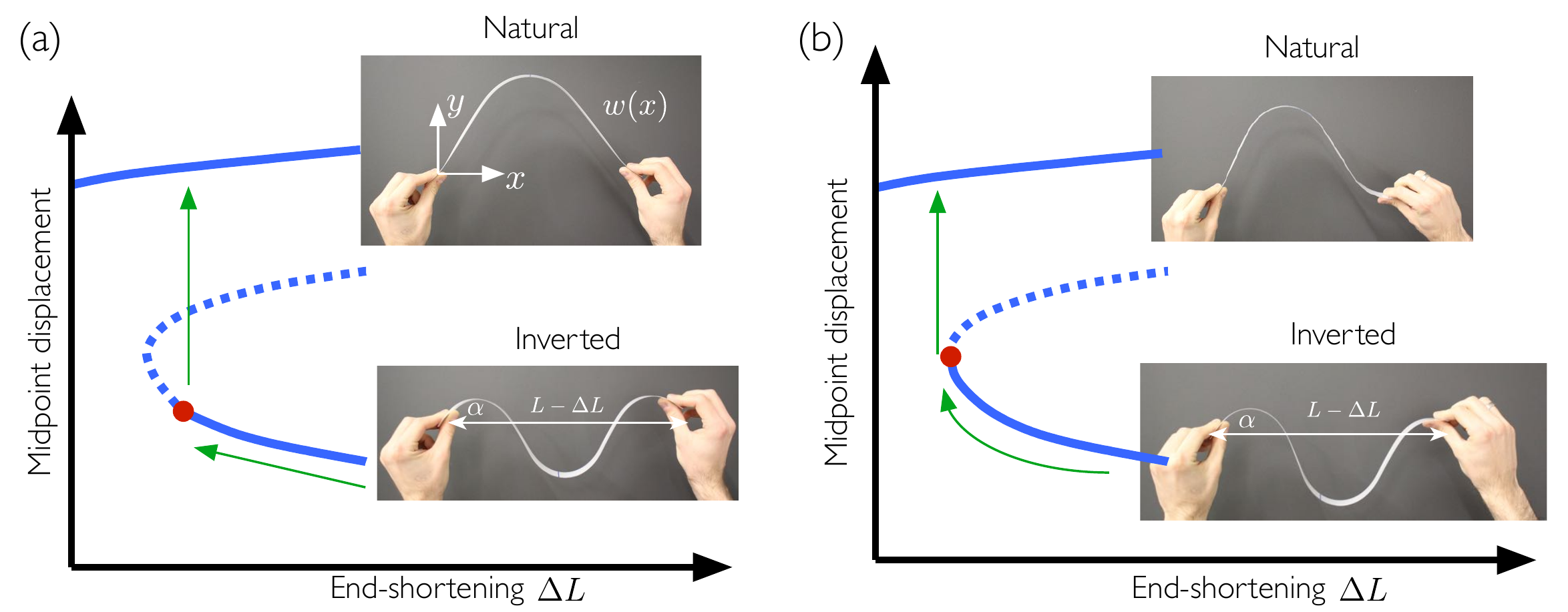}
\caption{Exploring `snap-through' instabilities in a simple elastic system. (a) Bringing the edges of a strip of plastic together, while also holding them at a non-zero angle $\alpha$ to the horizontal, creates bistable `inverted' (bottom) and `natural' (top) arch shapes. Under smaller end-shortenings $\Delta L$, the arch snaps from the inverted to the natural shape. Analysing the bifurcation behaviour shows that the instability underlying snapping in this case is a subcritical pitchfork bifurcation: the inverted mode (lower solid curve) intersects an unstable asymmetric mode (not drawn) at the point marked with a red dot; here it becomes linearly unstable (dashed curve). (Because the asymmetric mode is unstable it is not observed in practice.) (b) Introducing asymmetry in the boundary conditions, by holding the right end horizontally, still creates a bistable system. However, the destabilizing effect of the asymmetric mode is removed and the inverted mode remains stable up to a fold (indicated by the red dot): the snap-through bifurcation is now a saddle-node/fold bifurcation.  }
\label{arch}
\end{figure*}

Bistability and snap-through can be demonstrated using an elastic strip of length $L$ whose edges are clamped at equal angles $\alpha\neq0$ to the horizontal (Fig.~\ref{arch}a) \cite{Brinkmeyer2013,plaut_vibration_2009}. Provided that the two ends of this arch are brought together by a large enough distance $\dL$, two stable shapes exist: the `natural' mode and an `inverted' shape (Fig.~\ref{arch}a). However,  as $\dL$ is gradually decreased (pulling the two ends apart), the inverted equilibrium shape suddenly snaps through to the natural shape. The bifurcation diagram for this system (Fig.~\ref{arch}a) shows that as $\dL$ is decreased the inverted state becomes unstable, before ceasing to exist at still smaller $\dL$.

Snap-through due to an equilibrium state becoming unstable, as in the above example, is generally amenable to linear stability analysis \cite{pandey_dynamics_2014,fargette_elastocapillary_2014}: the displacement of each point on the arch evolves in time as $\sim e^{\sigma t}$ for some growth rate $\sigma$.  The more dynamically interesting  snap-through occurs when an equilibrium state ceases to exist without first becoming unstable (known as a saddle-node/fold bifurcation or limit-point instability \cite{Pippard1985}). This textbook snap-through \cite{Bazant} can be obtained simply by holding one end of the strip horizontally, while the other remains clamped at the angle $\alpha$ (Fig.~\ref{arch}b). 

The change in bifurcation to a saddle-node type means that a standard linear stability analysis no longer applies. Many previous works adopt a purely numerical approach to study the dynamics in this scenario \cite{Brinkmeyer2013}. In a simple elastic model, transverse displacements $w(x,t)$ of the strip are governed by the dynamic beam equation:
\beq
\rhos h\pdd{w}{t} + B\pdf{w}{x} +P\pdd{w}{x} = 0, \qquad 0 < x < L, \label{eqn:beamdim}
\eeq  with the material properties of the strip denoted by $\rhos$ (density), $h$ (thickness) and $B=Eh^3/12$ (bending stiffness, with $E$ the Young's modulus); $P$ is the applied compressive load (per unit width). An alternative approach is to estimate the time scale of snapping by balancing the first two terms in \eqref{eqn:beamdim}, suggesting that snap-through should occur on a time scale 
\beq
\tast=\left(\frac{\rhos h L^4}{B}\right)^{1/2}\sim \frac{L^2}{h\sqrt{E/\rhos}}.
\label{eqn:tsnap}
\eeq 
This time scale involves the speed of sound  within the strip, $(E/\rhos)^{1/2}$, and so is typically very short.  However, the above estimate frequently overestimates the speed of snapping, with the discrepancy attributed to some form of dissipation  \cite{Forterre2005,Brinkmeyer2013}. 
We investigate this snap-through `bottleneck' using a controlled version of our handheld snapping experiment and detailed analysis of \eqref{eqn:beamdim} . 

\begin{figure}
\centering
\includegraphics[width = 1\columnwidth]{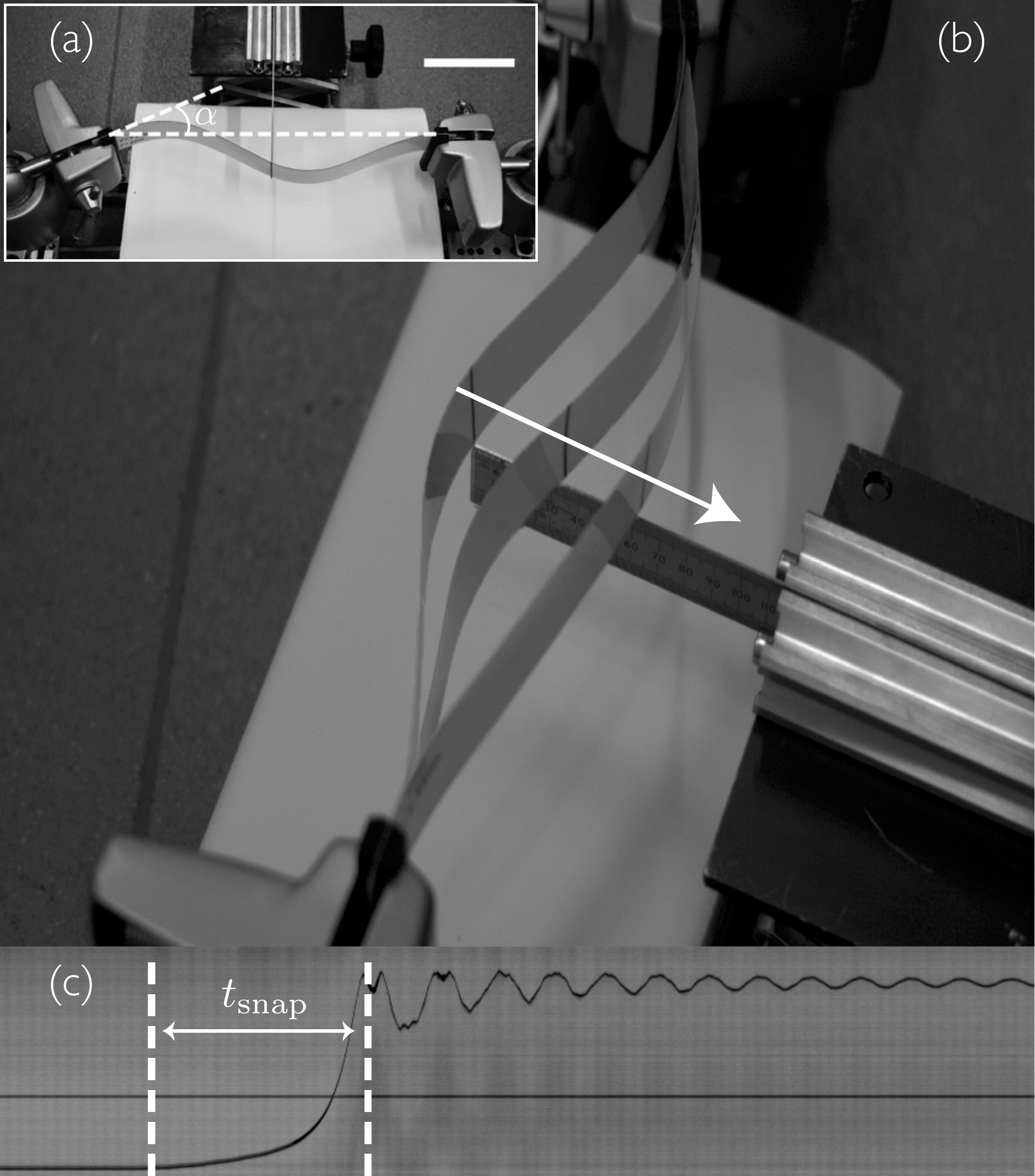}
\caption{ Investigating the snapping dynamics of an elastic arch. (a) A thin strip is buckled into an unstable state with an end-shortening past the snapping transition, $\Delta L < \Lf$; a metal indenter prevents the strip from snapping by making contact at its midpoint and imposes $w(L/2,0) = \wf(L/2)$.  Scale bar $10~\mathrm{cm}$. (b) The indenter is then lowered allowing the arch to snap (three successive stages superimposed). (See also Methods.) (c) A spatio-temporal plot of the midpoint reveals its trajectory during snapping (PET, $L = 240~\mathrm{mm}$, $\alpha = 21.34^{\circ}$, $\Lf = 10.41~\mathrm{mm}$, $\wf(L/2) = -16.75~\mathrm{mm}$, $\Delta L = 10.20~\mathrm{mm}$). The montage begins before the point when the indenter loses contact with the strip, and ends as the arch oscillates about the natural shape (the horizontal line is at zero displacement). Slices through a total of $828$ frames (separated by $1\mathrm{~ms}$) are shown.}
\label{setup}
\end{figure}

We performed experiments on thin strips of polyethylene terephthalate (PET) and stainless steel shim (see Methods). A strip is clamped (with ends angled appropriately) and then buckled into an arch by imposing an end-shortening $\Delta L$ (Fig.~\ref{setup}a). Snap-through is reached by altering $\dL$ quasi-statically to values $\dL<\Lf$, with the threshold $\Lf$ determined experimentally (see appended Supplementary Information for details). To obtain repeatable experiments with a given value of $\dL < \Lf$, the strip is initially prevented from snapping by an indenter that fixes the displacement of the midpoint to be that  at the bifurcation point,  $\wf(L/2)$. On removing the constraint, the strip then snaps from rest (Fig.~\ref{setup}b). 


\begin{figure}
\centering
\includegraphics[width = \columnwidth]{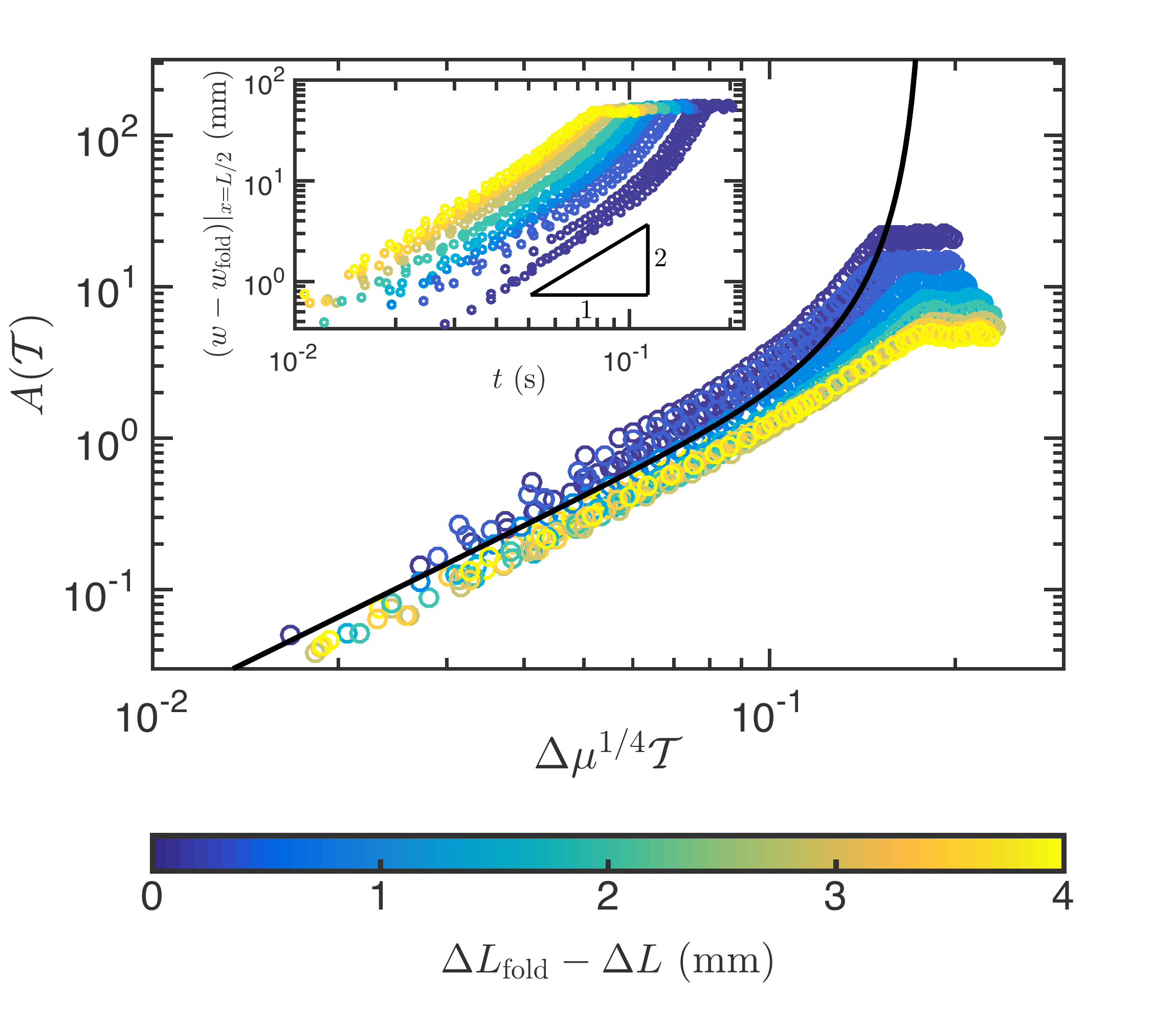} 
\caption{Midpoint trajectories during snapping for different end-shortenings just beyond the snapping transition. Inset: Evolution of the midpoint position, $w(L/2,t)$, away from the initial value $\wf(L/2)$ (PET, $L = 290~\mathrm{mm}$, $\alpha = 19.85^{\circ}$, $\Lf = 9.20~\mathrm{mm}$, $\wf(L/2) = -20.95~\mathrm{mm}$).  For each value of $\Lf - \Delta L$ (given by the colour bar), three runs are recorded and shown here (circles). The time origin, $t = 0$, is the point where contact is first lost with the indenter (see Supplementary Information for details);  data is  plotted until the strip begins to oscillate  about the natural shape. Main plot: The same data, rescaled in terms of the amplitude variable $A(\tnd)$ as a function of dimensionless time $\tnd=t/t^{*}$. We see that while $A(\tnd)$ remains small, the points collapse onto the predicted asymptotic behaviour \eqref{beam10} (solid black curve) with quadratic growth initially.}
\label{rawtraj}
\end{figure}

\begin{figure}
\centering
\includegraphics[width = \columnwidth]{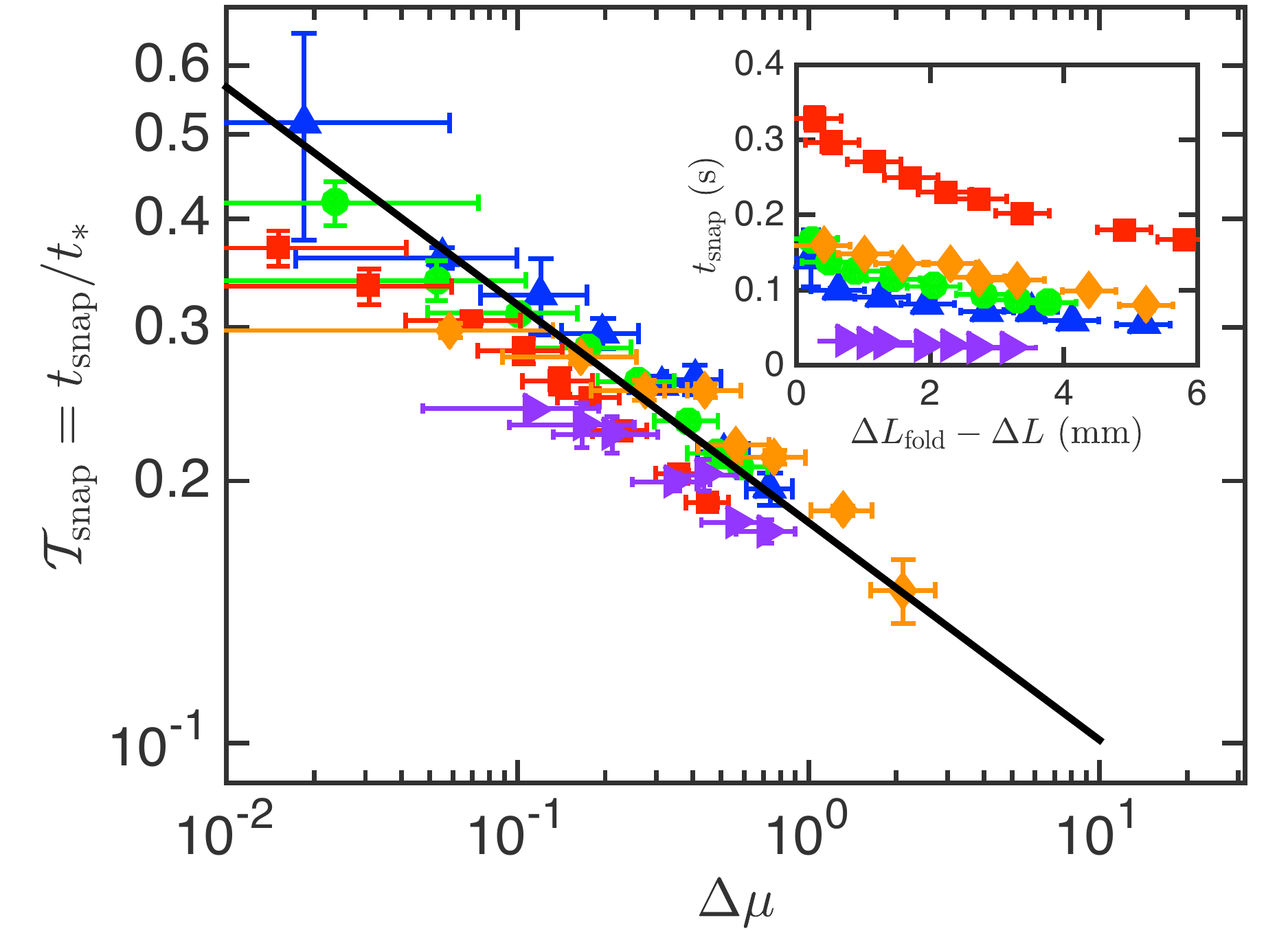}
\caption{Slowing down of snapping dynamics near the loss of bistability. Inset: The experimentally measured snapping time, $\tb$ (defined as the time taken to reach the first peak of vibrations, as labelled in Fig.~\ref{setup}c, averaged over three runs), as the end-shortening $\Delta L$ approaches the snapping threshold $\Lf$ from below. Data is shown for PET strips with $L = 240~\mathrm{mm}$, $\alpha =21.34^{\circ} $ (upward-pointing triangles), $L = 290~\mathrm{mm}$, $\alpha = 19.85^{\circ}$ (circles) and $L = 430~\mathrm{mm}$, $\alpha = 21.17^{\circ} $ (squares), as well as for experiments on steel strips with $L = 140~\mathrm{mm}$, $\alpha = 22.51^{\circ}$ (right-pointing triangles) and $280~\mathrm{mm}$, $\alpha = 17.14^{\circ} $ (diamonds). Main plot: Snapping times, re-scaled by the inertial timescale $t^{*}$, as a function of the normalized distance to bifurcation $\Dmu = \mu-\muf$ (computed from \eqref{eqn:defnmu}). The data collapse onto the asymptotic prediction $\tnd_{\mathrm{snap}}\approx 0.179\Dmu^{-1/4}$ from linear beam theory (solid black line). Horizontal error bars correspond to the uncertainties  in measurements of $\alpha$ ($\pm 2^{\circ}$) and $\Delta L$ ($\pm 200~\mu\mathrm{m}$)  (see Methods); vertical error bars correspond to the standard deviation of the measured snapping times over three runs. }
\label{rawtimes}
\end{figure}

A spatio-temporal plot of the midpoint position during snap-through (Fig.~\ref{setup}c) reveals the nonlinear nature of the early stages  and the under-damped oscillations about the natural state. The inset of Fig.~\ref{rawtraj} shows that the displacement of the midpoint initially grows quadratically in time, in contrast to the exponential growth that is  observed in systems at the onset of instability \cite{pandey_dynamics_2014,fargette_elastocapillary_2014};  a systematic slowing down is also seen in both the displacement and the snap-through time, $\tb$ (inset of Fig.~\ref{rawtimes}), as the bifurcation point $\Lf$ is approached.  This slowing down behaviour is reminiscent of critical slowing down in a number of physical phenomena \cite{Lubensky,Tredicce2004,Strogatz1989,Dakos2008,Scheffer2009}. 

To understand these observations, we analyse the linear beam equation, \eqref{eqn:beamdim}, incorporating the lateral confinement of the (inextensible) arch, which for small deflections is approximated by
\beq
\int_0^L\left(\pd{w}{x}\right)^2~\id x=2\dL.
\label{eqn:constraint}
\eeq The inclination angle $\alpha$ enters through the boundary condition $w_x(0,t)=\alpha$ (subscripts denote differentiation) while the other boundary conditions are homogeneous, $w(0,t)=w(L,t)=w_x(L,t)=0$. The linearity of the beam equation \eqref{eqn:beamdim} and the form of the confinement \eqref{eqn:constraint} show that $w_x$ can be rescaled by $(\dL/L)^{1/2}$ and, using $L$ as the natural horizontal length scale, we see immediately that a single dimensionless parameter emerges:
\beq
\mu = \alpha\left(\frac{\dL}{L}\right)^{-1/2}. \label{eqn:defnmu}
\eeq The purely geometrical parameter $\mu$ is the key control parameter in the problem and may be  understood as the ratio of the angle imposed by clamping, $\alpha$, to that imposed by confinement, $(\dL/L)^{1/2}$.
 
An analysis of the equilibrium solutions of \eqref{eqn:beamdim} subject to the constraint \eqref{eqn:constraint} and the appropriate boundary conditions allows the shape of the inverted arch to be determined analytically where it exists (see Supplementary Information for details). This analysis confirms that bistability of the inverted shape is lost at a saddle-node bifurcation where $\mu = \muf \approx 1.782$; the bifurcation shape, $\wf(x)$, and the associated (constant) compressive force, $P L^2/B=\tauf^2\approx 57.55$, can also be found explicitly. 

To analyse the dynamics just beyond the bifurcation point, i.e.~for end-shortenings $\Delta L$ slightly smaller than $\Lf$, we set $\mu = \muf + \Dmu$ with $0 < \Dmu \ll 1$. Because no inverted equilibrium exists for $\Dmu>0$, we exploit  the fact that the midpoint displacement is initially the same as at the bifurcation, i.e.~$w(L/2,0) = \wf(L/2)$. For small $\Dmu$, the initial shape of the strip is therefore `close' to the bifurcation shape everywhere, $w(x,0) \approx \wf(x)$ for $0 < x < L$. 
Introducing dimensionless variables $X=x/L$, $W = w/(L\dL)^{1/2}$, $\Wf = \wf/(L\dL)^{1/2}$, $\tnd=t/\tast$ and $\tau^2 = P L^2/B$ we then seek a series solution of the form
\begin{eqnarray}
W(X,\tnd) & = & \Wf(X) + \Dmu^{1/2} W_p(X)A(\tnd), \label{beam5}\\
\tau(\tnd) & = & \tauf + \Dmu^{1/2}A(\tnd).\label{beam6}
\end{eqnarray} We find that $W_p(X)$ is a spatial eigenfunction that can be determined analytically, while the temporal variation of the motion, $A(\tnd)$, satisfies
\beq
\Dmu^{-1/2} \dd{A}{\tnd} = c_1 + c_2 A^2, \label{beam9}
\eeq
with $c_1 \approx 329.0$ and $c_2 \approx 417.8$ constants that depend on $W_p(X)$ and its integrals (see Supplementary Information).

The ordinary differential equation \eqref{beam9} represents a great simplification of the full system \eqref{eqn:beamdim} and \eqref{eqn:constraint}. Furthermore, \eqref{beam9} is generic for the dynamics of elastic snap-through without dissipation: the nonlinear term $A^2$ is inherited from the structure of the saddle-node bifurcation (the parabolic geometry near the fold) and should hold generically in inertial snap-through. With viscous damping, a similar analysis would lead to a single time derivative in \eqref{beam9}. The specific details of the problem (e.g. boundary conditions) enter only through the  constants $c_1$ and $c_2$.

Because the arch starts from rest at a shape close to the bifurcation shape, the appropriate initial conditions for \eqref{beam9} are $A(0) = A_{\tnd}(0) = 0$, giving the implicit solution
\beq
\Dmu^{1/4} \tnd= \sqrt{\frac{3}{2}}\int_0^{A(\tnd)} \frac{\id \xi}{(3 c_1 \xi+ c_2 \xi^3)^{1/2}} . \label{beam10}
\eeq
Expanding the right-hand side for  $A\ll1$, we obtain the asymptotic behaviour 
\beqn
A(\tnd) \sim \frac{c_1}{2} \Dmu^{1/2} \tnd^2, 
\eeqn
when $\tnd \ll \Dmu^{-1/4}$. The initial growth of the midpoint displacement (and all other material points) is therefore ballistic, as observed experimentally (Fig.~\ref{rawtraj} inset); viscous dissipation would instead give $A\propto\tnd$. Furthermore, the full solution for $A(\tnd)$ computed from \eqref{beam10}  compares favourably to that observed experimentally while $A\lesssim1$ (Fig.~\ref{rawtraj}), with experimental data for different values of $\Dmu$ collapsing onto a single master curve.  For $A \gg 1$ our asymptotic analysis breaks down and fails to predict how the strip approaches the natural equilibrium and oscillates; instead, the solution \eqref{beam10} blows up at time $\tnd=\tbot$ where
\beq
 \tbot  = \sqrt{\frac{3}{2}}\Dmu^{-1/4}\int_0^{\infty} \frac{\id \xi}{\left(3 c_1 \xi+ c_2 \xi^3\right)^{1/2}}\approx 0.179 \Dmu^{-1/4}.
\label{beam11}
\eeq This time therefore corresponds to the end of the `bottleneck phase' in which the strip is influenced by its proximity to the inverted equilibrium at the fold. Because the motions are rapid by this stage, $\tbot$ represents a natural approximation for the total snap-through time. The prediction \eqref{beam11} leads to a collapse of experimentally measured snapping times for strips composed of different materials and natural lengths $L$, and predicts the dependence on $\Dmu$ well (Fig.~\ref{rawtimes}). 

The data in Fig.~\ref{rawtimes} and the expression for the snap-through time, \eqref{beam11}, show that as the system approaches the snap-through transition the dynamics slow down significantly. This is characteristic of the dynamics near a bifurcation in a range of physical phenomena \cite{Tredicce2004,Strogatz1989,Dakos2008,Scheffer2009,aranson_crystallization_2000}, which is commonly referred to as `critical slowing down' \cite{Lubensky}, or as a `bottleneck' due to the `ghost' of the nearby equilibrium \cite{Strogatz}.  The importance of critical slowing down in elastic instabilities such as snap-through has not been appreciated previously,  despite numerous previous experiments showing signs of diverging time scales as the threshold is approached \cite{Brinkmeyer2013,Hung1999}. Furthermore,  the inertial (rather than overdamped) dynamics here changes the exponents typically seen in critical slowing down ($\Dmu^{-1/4}$ rather than $\Dmu^{-1/2}$ \cite{aranson_crystallization_2000}).

We note that  the prefactor in the scaling $\tbot\sim\Dmu^{-1/4}$ depended on a detailed calculation and hence on the boundary conditions of the problem;  here the prefactor is small, meaning that the snap-through time is comparable to the characteristic elastic time scale $\tast$ for experimentally attainable $\Dmu$. However, in other systems the appropriate prefactor may be substantially larger and finer control of the distance to bifurcation $\Dmu$ may be possible; in such circumstances we expect that a substantial disparity between the observed snap-through time and the characteristic elastic time scale $\tast$ may emerge.  Biological systems such as the Venus flytrap may be particularly prone to such a slowing down, as the analogue of $\Dmu$ is often controlled by slow processes such as swelling or growth.  In both biological and engineering settings, snap-through would seem to require a trade-off between the speed of snapping (a faster snap requiring larger $\Dmu$) and the time/energy taken to attain a large $\Dmu$.  Critical slowing down may also mean that very close to the snap-through transition the system becomes overdamped (rather than inertial) leading to different scalings and slower dynamics; this possibility remains to be fully explored and may depend on the precise nature of the damping present (e.g.~viscoelasticity in man-made applications \cite{Brinkmeyer2013} or poroelasticity in biological systems \cite{Forterre2005,Skotheim2005}). Our analysis, combined with techniques for controlling snap-through such as solvent-induced swelling \cite{Holmes2007} or photo-initiation \cite{RaviShankar2013}, may offer the possibility to tune the time scale of snap-through from fast to slow by controlling how far beyond the transition one takes the system.


\acknowledgments

We are grateful to Jean-Baptiste Gorce for early experiments in a related system and to Jonathan Dawes, S\'{e}bastien Neukirch, and Jin-Chong Tan for discussions. The research leading to these results has received funding from  the European Research Council under the European Union's Horizon 2020 Programme / ERC Grant Agreement no.~637334 (DV) and the EPSRC Grant No.~EP/ M50659X/1 (MG). 

\section*{Supplementary information}
Further detail on the theoretical analysis of the snapping problem and the experimental procedures is available in the Supplementary Information (appended).

\section*{Data availability}
The experimental data that supports the plots within this paper and other findings of this study are available from http://dx.doi.org/10.5287/bodleian:RyGXnqJGk.

%
%

\section*{Methods}
\small{\textbf{Sample preparation.}~Strips were prepared from biaxially oriented polyethylene terephthalate (PET) film (Goodfellow, Cambridge, $\rho_s = 1.337\mathrm{~g\,cm^{-3}}$, $h = 0.35\mathrm{~mm}$, $E=5.707\mathrm{~GPa}$) and stainless steel rolled shim ($304$ grade, RS components, $\rho_s = 7.881\mathrm{~g\,cm^{-3}}$, $h = 0.1\mathrm{~mm}$, $E=203.8\mathrm{~GPa}$). The value of the Young's modulus $E$ for each material were determined  by analysing the frequency of small-amplitude vibrations of the strip. The time scale of snapping was varied by varying the length of the strip: for PET we used lengths $L \in\{240,290,430\} \mathrm{~mm}$ while for steel we used $L\in\{140,280\} \mathrm{~mm}$. Experiments were conducted at room temperature, well below the glass transition temperature of PET, so that this material acts as a glassy polymer, showing little viscous creep and dynamic dissipation (see \cite{Shi2001,Demirel2011} for example).

\small{\textbf{Snapping experiments.}~The ends of each strip are clamped into vice clamps (PanaVise $301$) which are mounted onto a linear track so that the strip deforms in one plane only. To minimize the effect of gravity, the strip is oriented sideways so its width lies in the vertical direction (see Fig.~2a,b of the main text). The right clamp is fixed parallel to the track, while the left clamp holds the strip at an angle $\alpha \neq 0$ (constant throughout each experiment) and can be moved along the track to vary the applied end-shortening $\Delta L$. A digital camera mounted above the left clamp allows $\alpha$ to be determined to an accuracy of $\pm 2^{\circ}$, and changes in $\Delta L$ to be measured to an accuracy of $\pm 200~\mu\mathrm{m}$ (by measuring displacement of the clamp from a known reference state).  

The snapping dynamics are filmed using a high speed camera (Phantom Miro $310$) at a frame rate of $1000 \mathrm{~fps}$. The camera is placed vertically above the strip, allowing the midpoint position (marked on the edge) to be recorded when the strip is in equilibrium and during motion. Beyond the fold, i.e.~when only one equilibrium exists, the strip is forced to start close to the fold shape using a metal ruler (tip width $1\mathrm{~mm}$) attached to a laboratory jack; this is then lowered vertically out of contact with the strip to allow snapping to proceed. The resulting movie is cropped around the midpoint position and converted to a spatio-temporal plot (montage) of its trajectory using ImageJ (NIH).}

\widetext
\clearpage
\begin{center}
\textbf{\large Supplementary information for ``\emph{Critical slowing down in purely elastic `snap-through' instabilities}"}
\end{center}
\setcounter{equation}{0}
\setcounter{figure}{0}
\setcounter{table}{0}
\setcounter{page}{1}
\makeatletter
\renewcommand{\theequation}{S\arabic{equation}}
\renewcommand{\thefigure}{S\arabic{figure}}
\renewcommand{\bibnumfmt}[1]{[S#1]}
\renewcommand{\citenumfont}[1]{S#1}

This supplementary information gives further detail on the theoretical analysis of the snapping problem and the experimental procedures than is possible in the main text. In particular, \S\ref{sec:theory} discusses the use of linear beam theory to characterize  the bistability of our system and the early-time dynamics of snap-through. In \S\ref{sec:expt} we discuss how we experimentally determine the location of the fold point and extract the midpoint trajectory from the movies of snapping. 

\section{Theoretical analysis\label{sec:theory}}
The transverse displacement of the strip, $w(x,t)$, is modelled using the linear dynamic beam equation \citep{howell} 
\beq
\rhos h\pdd{w}{t} + B\pdf{w}{x} +P\pdd{w}{x} = 0, \qquad 0 < x < L. 
\label{eqn:beamdimSI} 
\eeq
Here $x$ is the horizontal coordinate (measured from the left end), $t$ is time, and $P(t)$ is the (unknown) compressive force applied to the strip (per unit width). The properties of the strip are its natural length $L$, thickness $h$, density $\rho_s$, and bending stiffness $B = Eh^3/12$  (with $E$ the Young's modulus). (Note that we use the bending stiffness appropriate for a narrow strip, see \cite{audoly2010elasticity}.)

The left end of the strip is clamped at an inclination angle $0<\alpha\ll1$, while the right end is clamped horizontally. This corresponds to the boundary conditions (here and throughout subscripts denote partial derivatives) 
\beq
w(0,t) = 0, \quad w_x(0,t) = \alpha, \quad w(L,t)=w_x(L,t) = 0. \label{eqn:bcsdim}
\eeq

We consider slender strips, $h \ll L$, well beyond the buckling threshold so that the extensibility of the strip may be neglected (see \cite{pandey_dynamics_2014SI} for a discussion of this in a related problem). The lateral confinement of the strip then becomes $\int_0^L \cos\theta~\id s = L-\Delta L$ where $\theta(s,t)$ is the angle made by the strip to the horizontal, $s$ is the arclength and $\dL$ is the imposed end-shortening. Using the assumption of small slopes implicit in the beam equation, we identify $s \sim x$ and $\theta \sim w_x \ll1$ so this is approximated as
\beq
\int_0^L\left(\pd{w}{x}\right)^2~\id x=2\dL. \label{eqn:constraintdim}
\eeq 

Equation \eqref{eqn:beamdimSI} with boundary conditions \eqref{eqn:bcsdim}, the constraint \eqref{eqn:constraintdim} and initial conditions fully specify the problem.  We take as initial conditions zero initial velocity, $w_t(x,0)=0$, and initial shape $w(x,0)=w_0(x)$, where the function $w_0(x)$ shall be specified later.

\subsection{Non-dimensionalization}

To make the problem dimensionless, we scale the horizontal coordinate by the length $L$ of the strip, i.e~we set $X=x/L$. Balancing terms in the inextensibility constraint \eqref{eqn:constraintdim} shows that a typical slope $w_x \sim (\Delta L/L)^{1/2}$, giving the natural vertical length scale $w \sim (L\Delta L)^{1/2}$. We therefore introduce the dimensionless displacement $W = w/(L\dL)^{1/2}$. Time is scaled as $\tnd=t/\tast$ where $\tast = (\rho_s h L^4/B)^{1/2}$ is the inertial time scale (obtained by balancing inertial and bending forces in \eqref{eqn:beamdimSI}). Inserting these scalings into the beam equation \eqref{eqn:beamdimSI}, we obtain 
\beq
\pdd{W}{\tnd} + \pdf{W}{X} +\tau^2\pdd{W}{X} = 0, \qquad 0 < X < 1, \label{eqn:beamnondim}
\eeq
where $\tau(\tnd)^2 = P L^2/B$ is the dimensionless compressive force.

With this non-dimensionalization, the constraint \eqref{eqn:constraintdim} becomes
\beq
 \label{eqn:constraintnondim}\int_0^1\left(\pd{W}{X}\right)^2~\id X=2, 
\eeq
while the boundary conditions \eqref{eqn:bcsdim} are modified to 
\beq
W_X(0,\tnd) = \mu \equiv \alpha (\Delta L/L)^{-1/2}, \quad W(0,\tnd) = W(1,\tnd) = W_X(1,\tnd) = 0. \label{eqn:bcsnondim}
\eeq
Together with appropriate initial conditions for $W$ and $W_{\tnd}$, these equations provide a closed system to determine the profile $W(X,\tnd)$ and compressive force $\tau(\tnd)^2$. 

By non-dimensionalizing the problem we have reduced the control parameters $\alpha$ and $\Delta L$ to the single parameter $\mu$, which enters the problem as a normalized inclination angle (the dimensionless compressive force $\tau^2$ acts as a Lagrange multiplier and is determined as part of the solution). The parameter $\mu$ measures the ratio of the angle imposed by clamping, $\alpha$, to that due to the imposed end-shortening, $(\Delta L/L)^{1/2}$, and so is entirely geometric in nature; it is independent of the material parameters of the system, notably $\rho_s$ and $E$ (and also the thickness $h$). While the combination of parameters encapsulated in $\mu$ has been identified empirically before \citep{Brinkmeyer2014}, the analytical understanding here is, to our knowledge, new. Finally, we note that $\mu$ is the ratio of bending energy to stretching energy within the strip and, as such, is the analogue of the F\"{o}ppl-von-K\'{a}rm\'{a}n number $\gamma$ for shallow spherical caps \citep{brodland_deflection_1987}.

\subsection{Equilibrium shapes}

The solution of the static beam equation \eqref{eqn:beamnondim} subject to the boundary conditions \eqref{eqn:bcsnondim} is
\beq
W(X) = \mu \frac{\tau X (\cos\tau - 1) + \tau[\cos\tau(1-X)-\cos\tau]-\sin\tau X - \sin \tau(1-X) + \sin\tau}{\tau(2\cos\tau+\tau\sin\tau-2)}. \label{eqn:steadysoln}
\eeq To determine $\tau$ in terms of the control parameter $\mu$, we substitute \eqref{eqn:steadysoln} into the end-shortening constraint \eqref{eqn:constraintnondim} and rearrange to find
\beq
\mu^2 = \frac{8\tau (2\cos\tau +\tau\sin\tau-2)^2}{2\tau^3 - \tau^2 (\sin 2\tau +4\sin\tau)+4\tau(\cos\tau-\cos 2\tau) +2 (\sin 2\tau - 2 \sin\tau)}. \label{eqn:taueqn}
\eeq

For each value of $\mu$, the allowed values of $\tau(\mu)$ may be found numerically (e.g.~using the \textsc{matlab} routine \texttt{fsolve}). Because $\tau$ does not manifest itself experimentally, we plot the resulting bifurcation diagram in terms of  $W(1/2)$ and $\mu$, using
\beq
W(1/2) = \mu \frac{\tan(\tau/4)}{2\tau},\label{eqn:midptdisp}
\eeq which follows from \eqref{eqn:steadysoln}. This result is shown in figure \ref{fig:fvkresponse}, and confirms that the `inverted' equilibrium shape (the red branch in figure \ref{fig:fvkresponse}) undergoes a saddle-node (fold) bifurcation when
\beqn
\mu = \muf \approx 1.7818, \quad W(1/2) = \Wf(1/2) \approx -0.3476, \quad \tau = \tauf \approx 7.5864.
\eeqn
Note that it can be shown via standard techniques (see \cite{maddocks1987stability}, for example), that the  branches represented by solid curves in figure \ref{fig:fvkresponse} are stable, while that represented by a blue dashed curve is unstable.

If the clamp angle $\alpha$ is fixed and $\mu$ is varied by changing the end-shortening $\Delta L$ (as in our experiments), the snapping bifurcation occurs at 
\beqn
\Lf \approx 0.315 \alpha^2 L.
\eeqn

\begin{figure}
\centering
\includegraphics[scale=0.59]{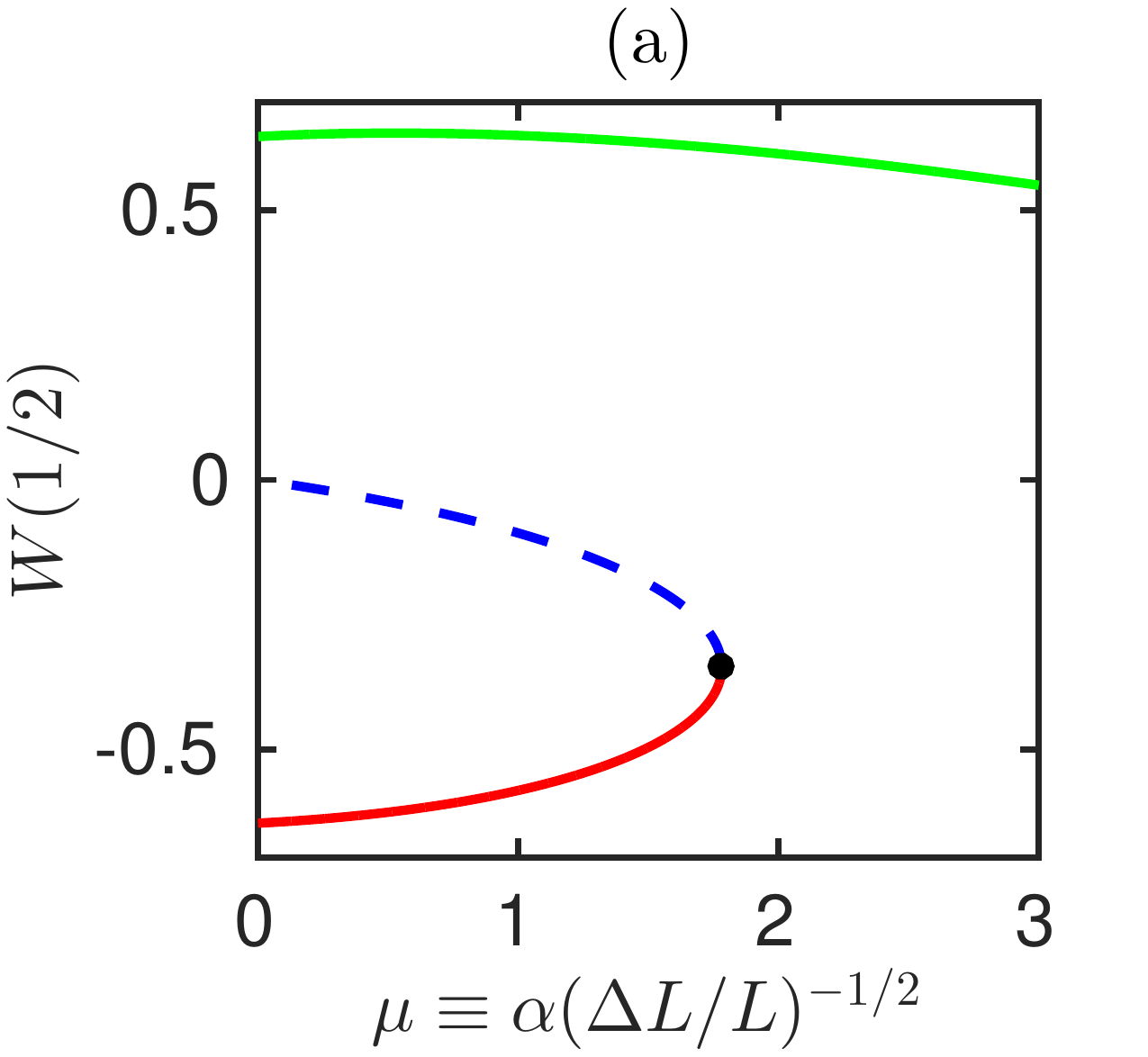} 
\includegraphics[scale=0.57]{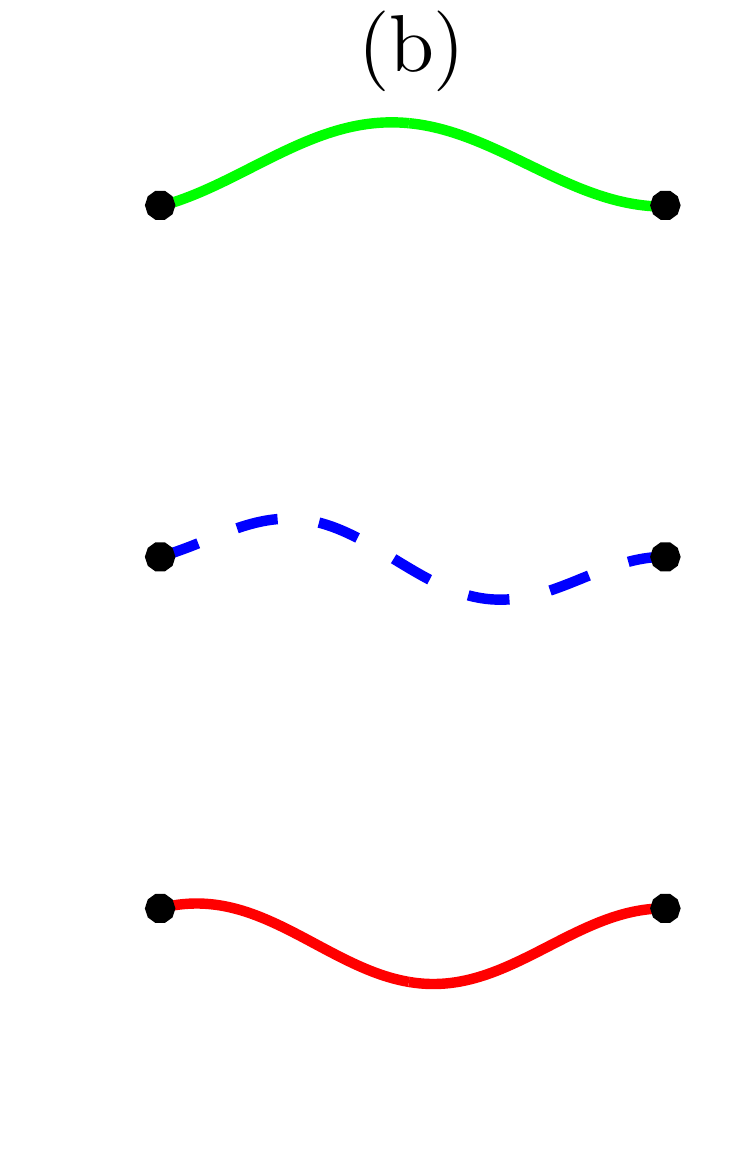} 
\caption{The equilibrium behaviour of the strip, as predicted by linear beam theory. (a) Bifurcation diagram: plotting the dimensionless midpoint displacement as a function of $\mu = \alpha (\Delta L/L)^{-1/2}$. The upper branch (solid green curve) corresponds to the `natural' shape, while the lower branch (solid red curve) corresponds to the `inverted' shape that disappears at the saddle-node (fold) bifurcation at $\mu\approx1.7818$. (b) The corresponding strip shapes, $W(X)$, for each of these modes when $\mu = 1$.}
\label{fig:fvkresponse}
\end{figure}

\subsection{Dynamics of snap-through}
We now analyse the situation in which the system is placed just beyond the saddle-node bifurcation, i.e.~we set $\mu = \muf + \Dmu$ with $0 < \Dmu \ll 1$, a small perturbation. Because no inverted equilibrium exists for $\Dmu > 0$ (i.e.~we are to the right of the fold point in figure \ref{fig:fvkresponse}), it follows that for the strip to reach an equilibrium, it must snap upwards to the natural shape. 

We take an initial condition that is `close' to the shape of the arch at the fold bifurcation, i.e.~$W(X,0)\approx\Wf(X)$. (Experimentally this condition is imposed by the indenter, which fixes the midpoint position to be that taken at bifurcation; away from the midpoint $W(X,0)\approx\Wf(X)$ then holds because $\Delta L \approx \Lf$ for $\Dmu \ll 1$). For $\Dmu$ sufficiently small,  we expect that the shape of the snapping beam will only evolve away from $\Wf(X)$ on a time scale that is much slower than the inertial time scale of the beam --- we have that $W(X,\tnd) \approx \Wf(X)$ (and also $\tau \approx \tauf$) up to some dimensionless time $\tnd \gg 1$. To capture this explicitly we rescale time as $T=\Dmu^\eta\tnd$, where $\eta>0$ characterizes the duration of this slow `bottleneck' phase and will be determined as part of the analysis. In terms of scaled variables, the beam equation \eqref{eqn:beamnondim} reads
\beq
\Dmu^{2\eta}\pdd{W}{T} + \pdf{W}{X} +\tau^2\pdd{W}{X} = 0, \qquad 0 < X < 1. 
\label{eqn:beamdimless} 
\eeq
We now seek a series solution about the bifurcation shape of the form
\begin{eqnarray}
W(X,T) & = & \Wf(X) + \Dmu^{1/2} \Wt_0(X,T)+\Dmu \Wt_1(X,T) + \ldots, \label{eqn:seriesW} \\
\tau(T) & = & \tauf + \Dmu^{1/2}\taut_0(T) + \Dmu\taut_1(T)+ \ldots. \label{eqn:seriestau} 
\end{eqnarray}
The choice of powers of $\Dmu^{1/2}$ can be justified \textit{a posteriori} and reflects the fact that the displacement (away from the bifurcation shape) in the bottleneck is much larger than the original perturbation to the system, $\Dmu$. A similar scaling also arises in bottleneck phenomena near the pattern forming threshold in the Swift-Hohenberg equation \citep{aranson_crystallization_2000SI}.

\subsubsection{Leading-order problem}

Inserting \eqref{eqn:seriesW}--\eqref{eqn:seriestau} into \eqref{eqn:beamdimless} and considering terms of $O(\Dmu^{1/2})$, we obtain the homogeneous equation
\beq
L(\Wt_0,\taut_0) \equiv \pdf{\Wt_0}{X} + \tauf^2 \pdd{\Wt_0}{X}+2\tauf \taut_0 \dd{\Wf}{X} = 0.  \label{eqn:leadingproblem}
\eeq
The end-shortening constraint \eqref{eqn:constraintnondim} and boundary conditions \eqref{eqn:bcsnondim} are also homogeneous at $O(\Dmu^{1/2})$:  
\beqn
\int_0^1 \d{\Wf}{X}\pd{\Wt_0}{X} \id X = 0, \quad  \Wt_0(0,T) = \Wt_{0_X}(0,T) = \Wt_0(1,T) = \Wt_{0_X}(1,T) = 0.
\eeqn
Because the leading-order problem is homogeneous, it is equivalent to the equations governing small-amplitude oscillations about the bifurcation shape $(\Wf,\tauf)$ restricted to neutrally stable (`slow') eigenmodes whose natural frequency (eigenvalue) is zero. Using linearity of the operator $L(\cdot,\cdot)$, we may scale $\taut_0$ out from \eqref{eqn:leadingproblem} (since $\taut_0$ is independent of $X$) so that
\beq
(\Wt_0,\taut_0) = A(T) (W_p(X),1). \label{eqn:leadingsoln}
\eeq
Here $A(T)$ is an (undetermined) amplitude and $W_p(X)$ is the eigenmode satisfying equation \eqref{eqn:leadingproblem} with $\taut_0 = 1$, i.e.~
\begin{eqnarray}
&& L(W_p,1) = 0, \quad \int_0^1 \d{\Wf}{X}\d{W_p}{X} \id X = 0, \label{eqn:Wpeqn} \\ 
&& W_p(0) = \left.\d{W_p}{X}\right|_{X=0} = W_p(1) = \left.\d{W_p}{X}\right|_{X=1} = 0. \label{eqn:Wpbcs}
\end{eqnarray}
While this system appears to over-determine $W_p(X)$ (there are four derivatives but five constraints), there is in fact a unique solution
\beq
W_p(X) = \frac{1}{\tauf}\left(X \d{\Wf}{X}-\muf X\right) + a_1\left(\sin\tauf X-\tauf X\right) + a_2\left(\cos\tauf X -1 \right),
\label{eqn:WpsolnExact}
\eeq where
\beqn
a_1 = -2 \muf \frac{\sin^2(\tauf/2)\left[(\tauf^2-2)\cos\tauf-2\tauf\sin\tauf+2\right]}{\tauf^2 \left(2\cos\tauf+\tauf\sin\tauf-2\right)^2}, 
\eeqn
and 
\beqn
a_2 = -\muf \frac{\tauf^3 + \tauf^2 \sin\tauf(\cos\tauf-2) + 2(\tauf\cos\tauf-\sin\tauf)(\cos\tauf-1)}{\tauf^2 \left(2\cos\tauf+\tauf\sin\tauf-2\right)^2}. 
\eeqn
(This solution is found by applying the boundary conditions \eqref{eqn:Wpbcs}, but also satisfies \eqref{eqn:Wpeqn} since, by construction, $\tauf$ is the value of $\tau$ taken at the fold.) 

Subsequently, we shall need two integrals associated with \eqref{eqn:WpsolnExact}; we record the values of these integrals here:
\beq
I_1=\int_0^1 W_p^2~\id X \approx 0.0518, \quad I_2=\int_0^1 \left(\d{W_p}{X}\right)^2 \id X \approx 0.950.
\label{eqn:Wpints}
\eeq

The variable $A(T)$ appearing in \eqref{eqn:leadingsoln} plays a key role in the snapping dynamics, because it acts as an amplitude of the leading-order solution (and also its compressive force). More explicitly, re-arranging the original series expansion in \eqref{eqn:seriesW} shows that
\beqn
A(T) \sim \Dmu^{-1/2}\frac{W(X,T)-\Wf(X)}{W_p(X)}.
\eeqn
The amplitude $A(T)$ therefore characterizes how the strip evolves away from the bifurcation shape during the bottleneck phase. Currently, $A(T)$ is undetermined. We proceed to the next order problem to determine $A(T)$.

\subsubsection{First-order problem}
The amplitude $A(T)$ will be determined by a solvability condition on the first-order problem.  In order to obtain dynamics at leading order, i.e.~for $A$ to be a non-constant function of time, the inertia term must come into play at $O(\Dmu)$, which requires $\eta=1/4$.  With this choice, at $O(\Dmu)$ the beam equation \eqref{eqn:beamdimless} becomes 
\beq
L(\Wt_1,\taut_1)  = -W_p \dd{A}{T}-A^2 \left(2\tauf \dd{W_p}{X} + \dd{\Wf}{X}\right). \label{eqn:firstproblem}
\eeq
The end-shortening constraint \eqref{eqn:constraintnondim} and boundary conditions \eqref{eqn:bcsnondim} at $O(\Dmu)$ have the form
\begin{eqnarray*}
&& \int_0^1 \d{\Wf}{X}\pd{\Wt_1}{X} \id X = -\tfrac{1}{2}A^2 \int_0^1 \left(\d{W_p}{X}\right)^2 \id X=-\tfrac{1}{2}I_2A^2, \\
&& \left.\d{\Wt_1}{X}\right|_{X=0} = 1, \quad \Wt_1(0,T) = \Wt_1(1,T) = \left.\d{\Wt_1}{X}\right|_{X=1} = 0.
\end{eqnarray*}

\subsubsection{Solution for $A(T)$}

Equation \eqref{eqn:firstproblem} features the same linear operator $L(\cdot,\cdot)$ as in the leading-order problem, but now with an inhomogeneous right-hand side. The Fredholm Alternative Theorem \citep{keener_principles_1988} implies that solutions  exist only for a certain right-hand side, yielding a solvability condition that takes the form of an ODE for $A(T)$. We formulate this condition in the usual way: we multiply \eqref{eqn:firstproblem} by the solution of the homogeneous adjoint problem, integrate over the domain, and use integration by parts to shift the operator onto the adjoint solution. In this case, the operator $L(\cdot,\cdot)$ is self-adjoint and so a solution of the homogeneous adjoint problem is simply $W_p(X)$. Performing these steps and simplifying (making use of the various boundary conditions that $\Wf$, $W_p$ and $\Wt_1$ satisfy) leads to 
\beq
\dd{A}{T} = c_1 + c_2 A^2, \label{eqn:Aeqn}
\eeq
where 
\beq 
c_1 = \frac{4 \tauf}{\muf I_1} \approx 329.0, \quad c_2 = \frac{3 \tauf I_2}{I_1} \approx 417.8,\label{eqn:c1c2}
\eeq
where the integrals $I_1$ and $I_2$ are as defined in \eqref{eqn:Wpints}. (Note that due to the scaled time $T=\Dmu^{1/4}\tnd$, this differs slightly to the amplitude equation (7) given in the main text.)

Similar analyses have previously been performed for over-damped dynamics in the Swift--Hohenberg equation \citep{aranson_crystallization_2000SI}, leading to the over-damped version of \eqref{eqn:Aeqn}, i.e.~with the second derivative on the left-hand side replaced by a first derivative. The overdamped version of \eqref{eqn:Aeqn} is well known as the canonical form of the dynamics close to a saddle-node bifurcation \citep{StrogatzSI} and is, in turn, known to give rise to a bottleneck whose duration diverges as the bifurcation point is approached. We shall see that \eqref{eqn:Aeqn} exhibits a similar bottleneck phenomenon but emphasize that the dynamics here is purely inertial: there is no dissipation in our model.

To solve \eqref{eqn:Aeqn}, we need initial conditions for $A$ and $A_{T}$. Because $\mu$ is perturbed by an amount $\Dmu$ from the fold point, it follows that $W(X,0)$ (the initial shape of the strip under the indenter) agrees with $\Wf(X)$ to within $O(\Dmu)$. (Alternatively, the steady indentation problem can be solved to obtain the shape of the beam, $W_{\mathrm{indent}}(X)$; comparing this with the bifurcation shape confirms that $|W_{\mathrm{indent}}(X)-\Wf(X)| = O(\Dmu)$.) The leading-order part enters the series solution in  \eqref{eqn:seriesW} at $O(\Dmu^{1/2})$ ($\gg \Dmu$), so that the initial conditions on $A$  are homogeneous, i.e.~
\beq
A(0) = A_T(0) = 0. \label{eqn:Adata}
\eeq 

The scaling $T=\Dmu^{1/4}\tnd$ confirms the assumption that the bottleneck regime extends up to $\tnd \gg 1$. As \eqref{eqn:Aeqn} is an $O(1)$ equation for $A(T)$ (ignoring numerical factors), the bottleneck time scales as
\beqn
\tnd \sim \Dmu^{-1/4} \gg 1.
\eeqn
In fact, we can evaluate the bottleneck time directly. Multiplying \eqref{eqn:Aeqn} by $A_{T}$ and integrating twice (taking the positive root for $A_{T}$) we obtain the full solution implicitly:
\beq
T = \sqrt{\frac{3}{2}} \int_0^{A(T)} \frac{\id \xi}{\left(3 c_1 \xi + c_2 \xi^3\right)^{1/2}}. \label{eqn:Asoln}
\eeq
As well as showing that $A(T)$ grows quadratically at early times, this solution also predicts that $A\to\infty$ in a finite time
\beqn
T_b  = \sqrt{\frac{3}{2}} \int_0^{\infty} \frac{\id \xi}{\left(3 c_1 \xi+ c_2 \xi^3\right)^{1/2}}  = \left(\frac{64 \pi^2 c_1 c_2 }{3}\right)^{-1/4} \Gamma\left(\tfrac{1}{4}\right)^2 \approx 0.179,
\eeqn
or, expressed in the dimensionless but unscaled time $\tnd$, $$\tnd_b \approx 0.179\Dmu^{-1/4}.$$ 

This blow-up corresponds to the end of the bottleneck regime --- the strip is no longer influenced by its proximity to the bifurcation shape, and is rapidly accelerating upward to the natural mode. Expanding the right-hand side of \eqref{eqn:Asoln} shows that the strip accelerates out of the bottleneck according to the power law
\beqn
A(T) \sim \frac{6}{c_2} (T_b-T)^{-2}.
\eeqn 
Returning to our original series \eqref{eqn:seriesW}--\eqref{eqn:seriestau}, we see that the perturbation variables are no longer small as soon as $A(T)$ grows to $O(\Dmu^{-1/2})$ and hence the amplitude equation \eqref{eqn:Aeqn} is no longer asymptotically valid. Nevertheless, our treatment here allows us to obtain the key quantity of interest, the snapping time, which is dominated by the time spent getting through the bottleneck.

\section{Experimental methods\label{sec:expt}}

\subsection{Determining the snapping transition}
The strip is first placed in the inverted equilibrium with a large enough end-shortening, $\Delta L$, so that the system is bistable. We then decrease $\Delta L$ quasi-statically in small steps, measuring the midpoint displacement, $w(L/2)$, until the strip snaps. The dimensional bifurcation diagram obtained in this way is shown in figure \ref{fig:loadtofold}a. 

\begin{figure}
\centering
\includegraphics[width =0.75\columnwidth]{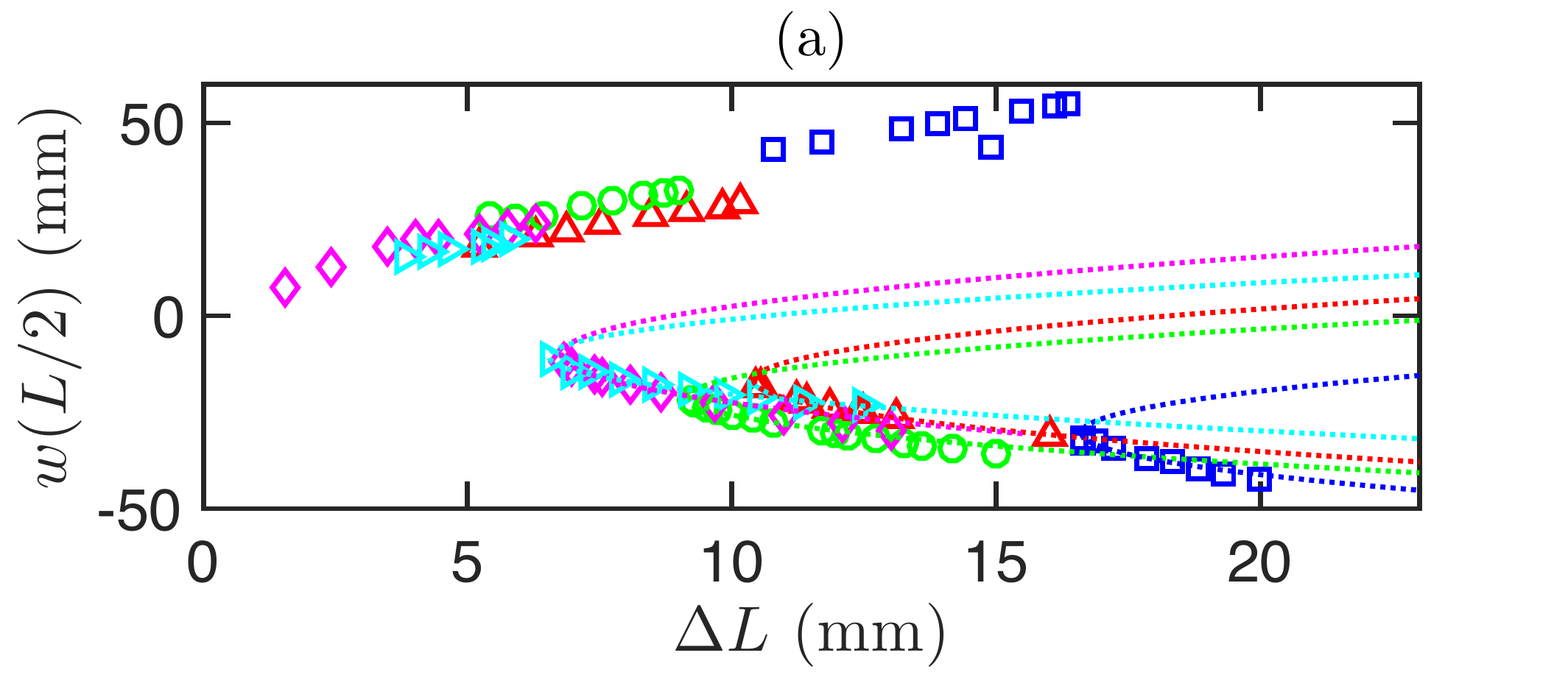} 
\includegraphics[width =0.75\columnwidth]{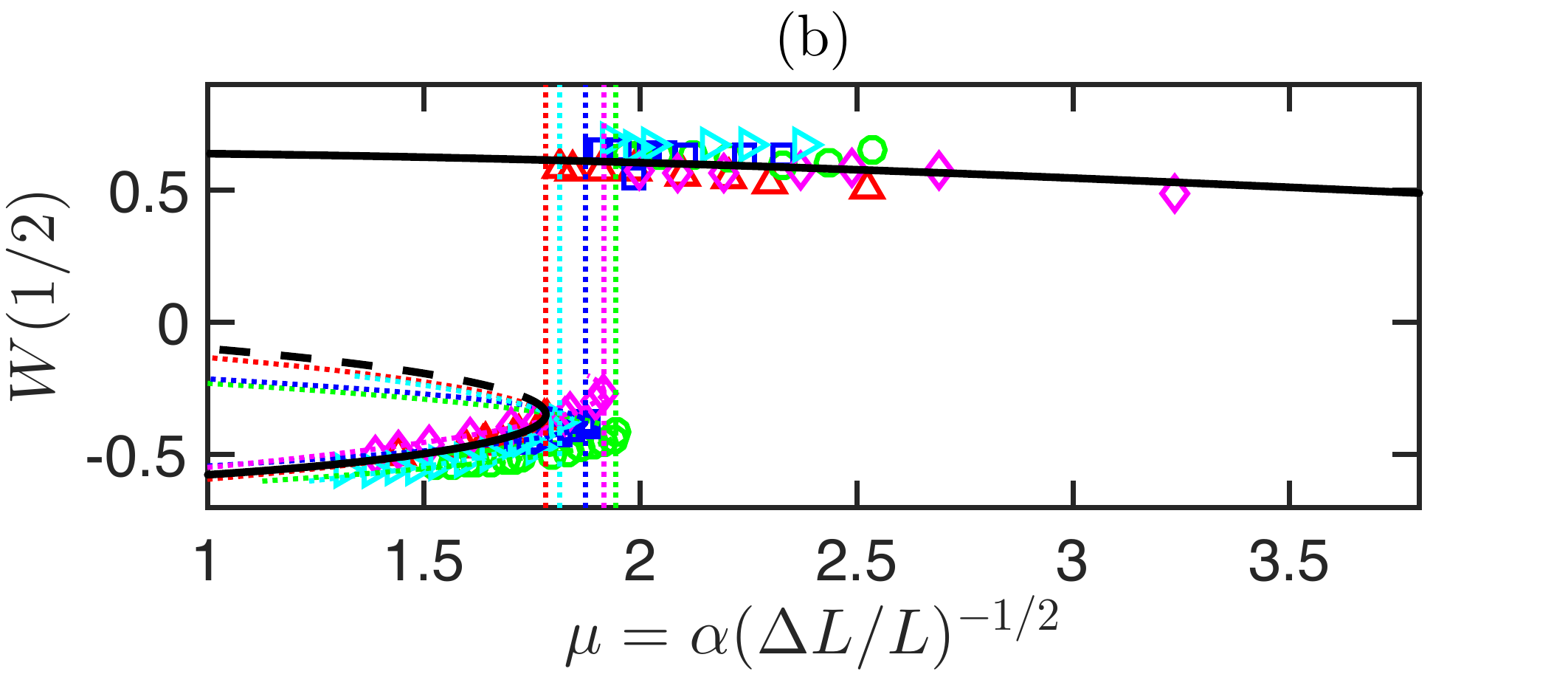} 
\caption{(a) Midpoint position of equilibrium shapes as a function of the applied end-shortening, $\Delta L$. Data is shown for PET strips with $L = 240~\mathrm{mm}$ (red upward-pointing triangles), $L = 290~\mathrm{mm}$ (green circles) and $L = 430~\mathrm{mm}$ (blue squares). Also plotted is data for steel strips with $L = 140~\mathrm{mm}$ (cyan right-pointing triangles) and $L = 280~\mathrm{mm}$ (magenta diamonds). The lower branches (i.e.~with $w(L/2) < 0$) correspond to the `inverted' shape while the upper branches (i.e.~with $w(L/2) > 0$) correspond to the `natural' shape after snap-through has occurred. In each case the best-fit (least-squares) parabola through the $6$ points closest to the snapping transition is shown (dotted curves). (b) The same data plotted in dimensionless terms, where $\alpha$ has been chosen within the range of experimental uncertainty ($\pm 2^{\circ}$) so that the fitted bifurcation points (vertical dotted lines) are close to the theoretical value  $\muf \approx 1.7818$. The final points are  $\muf \approx 1.7884$, $\alpha = 21.34^{\circ}$ (red upward-pointing triangles), $\muf \approx 1.9452$, $\alpha = 19.85^{\circ}$ (green circles), $\muf \approx 1.8800$, $\alpha = 21.17^{\circ}$ (blue squares), $\muf \approx 1.8174$, $\alpha = 22.51^{\circ}$ (cyan right-pointing triangles), and $\muf \approx 1.9236$, $\alpha = 17.14^{\circ}$ (magenta diamonds). The prediction from linear beam theory (reproduced from figure \ref{fig:fvkresponse}a) is also shown (solid black curves). }
\label{fig:loadtofold}
\end{figure}

The bifurcation point, $\Lf$, is never observed exactly: when we decrease $\Delta L$, we introduce perturbations that cause snap-through slightly before the fold is reached. We determine $\Lf$ by fitting the data points close to the transition to a parabola (dotted curves in figure \ref{fig:loadtofold}a) --- the generic form expected close to the fold. This best-fit parabola also predicts the corresponding midpoint position at the bifurcation point, $\wf(L/2)$, which is then the midpoint displacement fixed by the indenter. Note that this fitting procedure only needs to be performed once for each strip. 

In determining the bifurcation diagram experimentally, various errors are introduced in the measurement of the angle $\alpha$ and the end-shortening $\dL$. To plot the dimensionless bifurcation diagram (figure \ref{fig:loadtofold}b) we therefore allow the value of $\alpha$ to vary (within the limits of experimental uncertainty, $\pm2^\circ$) so that the experimentally determined fold point  is as close as possible to that predicted theoretically. The result, plotted in terms of $\mu = \alpha (\Delta L/L)^{-1/2}$,  is compared with the result of the linear beam theory in  figure \ref{fig:loadtofold}b. The fitted position of the fold, $\muf$ (as predicted from the best-fit parabola), is  displayed in each case (dotted vertical lines).  Because the snapping dynamics depend sensitively on the size of $\Dmu = \mu - \muf$, the theoretical value of $\muf$ ($\approx 1.7818$) is not used to calculate $\Dmu$; instead, we use the shifted values of $\alpha$, and the corresponding fitted values of $\muf$, since this is consistent with the observed behaviour of the strip prior to snapping. From here on, and throughout the main text, we refer to the shifted value of $\alpha$ for each experiment.

\subsection{Extracting the midpoint trajectory}

For values $\Delta L < \Lf$ (or equivalently $\mu > \muf$), the strip snaps upward to the natural shape once contact with the indenter is lost. Each snapping movie begins before the moment at which the indenter loses contact; after converting this to a spatio-temporal plot (montage) of the midpoint position, the trajectory is therefore initially flat on the montage and it is not clear from these plots \emph{when} the snapping motion first occurs. This start of the motion is key to the snapping time, and so this point must be determined by a fitting procedure. 


\begin{figure}
\centering
\includegraphics[width = 0.75\textwidth]{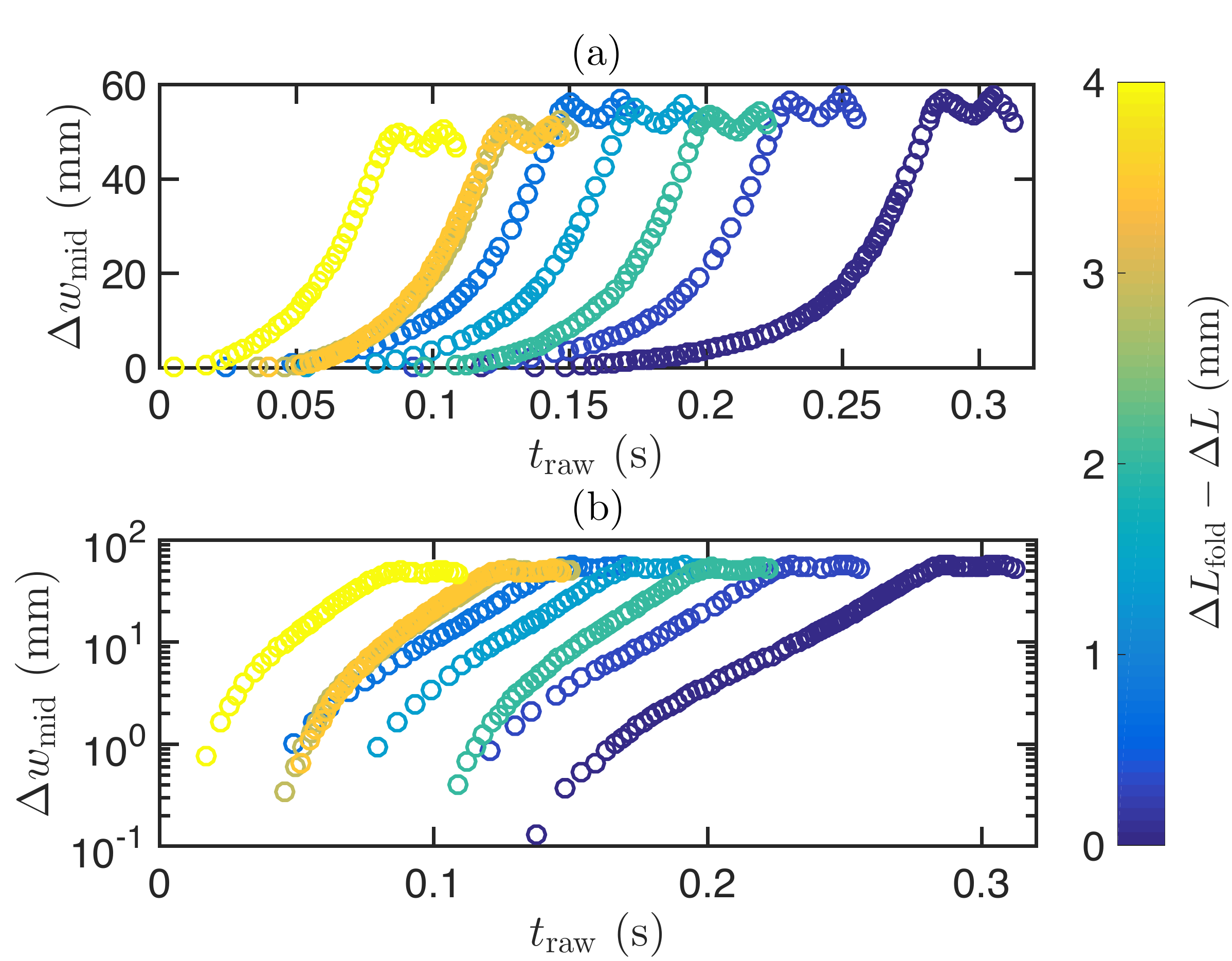}  
\caption{Midpoint trajectories during snap-through measured at different end-shortenings beyond the snapping transition (PET, $L = 290~\mathrm{mm}$, $\alpha = 19.85^{\circ}$, $\Lf = 9.20~\mathrm{mm}$, $\wf(L/2) = -20.95~\mathrm{mm}$). (a) Evolution of the midpoint displacement from its initial value, $\Delta w_{\mathrm{mid}} = (w - \wf)|_{x = L/2}$. For each end-shortening (given by colour bar), the snapping motion begins at some time $\tr=t_0 > 0$ that is not known. (Note that data is plotted only until the strip begins to oscillate about the natural shape). (b) The same data plotted on semi--log axes do not suggest the presence of a classical linear instability in the initial motion (which would be indicated by a phase of exponential growth).}
\label{fig:linearvssemilog}
\end{figure}

The values of $w(L/2,\tr)$ are determined from the montage, with $\tr$ denoting the raw time (measured from the arbitrary start of the montage). Figure \ref{fig:linearvssemilog}a shows a typical set of trajectories obtained in this way for snap-through at different values of $\Delta L < \Lf$; these are plotted in terms of the change in midpoint position away from the value imposed by the indenter, $\Delta w_{\mathrm{mid}} \equiv (w - \wf)|_{x = L/2}$.  Even without the moment of release determined, a semi--log plot of $\Delta w_{\mathrm{mid}}$ would reveal a straight line if the initial growth of the instability were exponential (as predicted by conventional stability analysis). This plot (figure \ref{fig:linearvssemilog}b) does not indicate such an exponential growth.

\begin{figure}
\centering
\includegraphics[width = 0.75\textwidth]{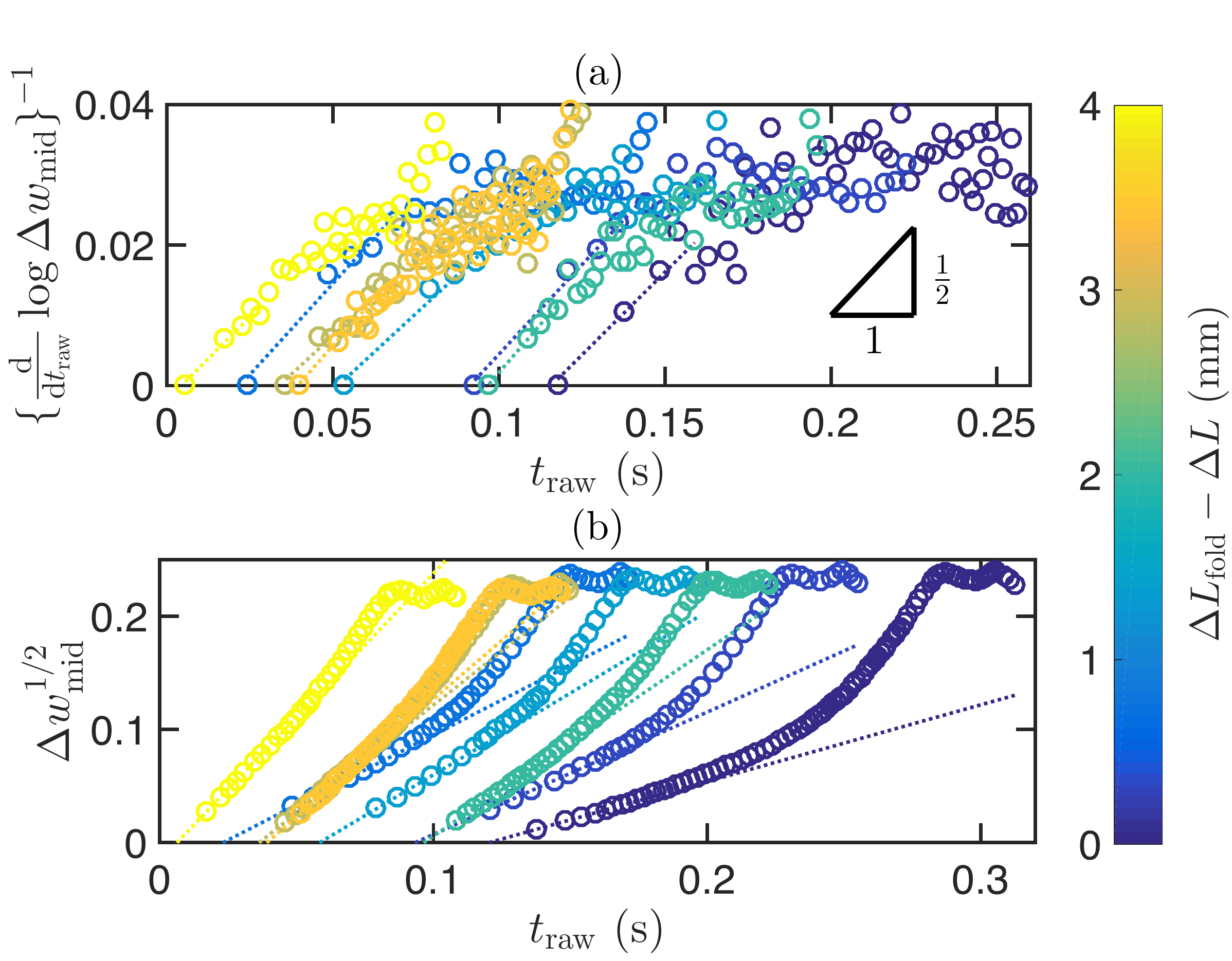}  
\caption{(a) The same data in figure \ref{fig:linearvssemilog} rescaled according to the right-hand side of \eqref{eqn:ansatz} (numerical differentiation was performed using forward differences). A linear (least-squares) fit over the first $5$ points in each case (dotted lines) gives the estimates $\beta \approx \left\lbrace 2.12,1.96,1.93,2.29,1.83,2.29,1.87,2.10\right\rbrace$ (given in increasing order of $\Delta L_{\mathrm{fold}}-\Delta L$); these are all consistent with the expected quadratic growth, $\beta=2$. (b) Rescaling the data according to the power-law \eqref{eqn:powerlaw} with $\beta = 2$, together with the best fit line over the first $5$ points in each case (dotted lines). This allows the start time of snapping, $t_0$, to be determined from the intercept with the horizontal axis.}
\label{fig:fittingstart}
\end{figure}

If we instead assume power-law growth of the form
\beq
\Delta w_{\mathrm{mid}} \propto (\tr - t_0)^{\beta}, \label{eqn:powerlaw}
\eeq
where $t_0$ is the time when contact is first lost and $\beta$ is an (unknown) exponent (assumed positive), then
\beq 
\frac{\tr - t_0}{\beta} = \left\lbrace \frac{\id}{\id \tr}\log\Delta w_{\mathrm{mid}}\right\rbrace^{-1}. \label{eqn:ansatz}
\eeq A plot of the experimentally determined RHS of \eqref{eqn:ansatz} as a linear function of $\tr$ is shown in figure \ref{fig:fittingstart}a. Despite the noise in the plot (which is due to numerical differentiation of the logarithm of our experimental data, as in \eqref{eqn:ansatz}), we see that our data is entirely consistent with $\beta = 2$, i.e.~the strip deflection grows quadratically in time. 

A further check that the growth is quadratic is to plot the square root of the midpoint displacement as a function of $\tr$ on linear axes (figure \ref{fig:fittingstart}b). This confirms a linear behaviour at early times, and allows us to determine the start of the snap, $t_0$, from the intercept of  the best-fit line with the horizontal axis. (We use the plotting procedure indicated in figure \ref{fig:fittingstart}b to determine $t_0$ because this is less susceptible to noise  than the approach used in figure \ref{fig:fittingstart}a.)

In the main text, the dimensionless midpoint trajectories are plotted in terms of the amplitude $A(\tnd)$, which is defined in the main text and implicitly via \eqref{eqn:seriesW} and \eqref{eqn:leadingsoln} to be
\beq
A(\tnd)=\frac{W(1/2,\tnd)-\Wf(1/2)}{\Dmu^{1/2}W_p(1/2)}.
\eeq In determining $A$, we use the value $W_p(1/2)\approx 0.3324$ determined from \eqref{eqn:WpsolnExact}. Figure \ref{fig:othercollapse} shows the experimentally determined $A(\tnd)$ that were not shown in figure $3$ of the main text; these experiments were performed with other lengths $L$ and strips made of either PET or steel.

\begin{figure}
\centering
\includegraphics[width = 0.95\textwidth]{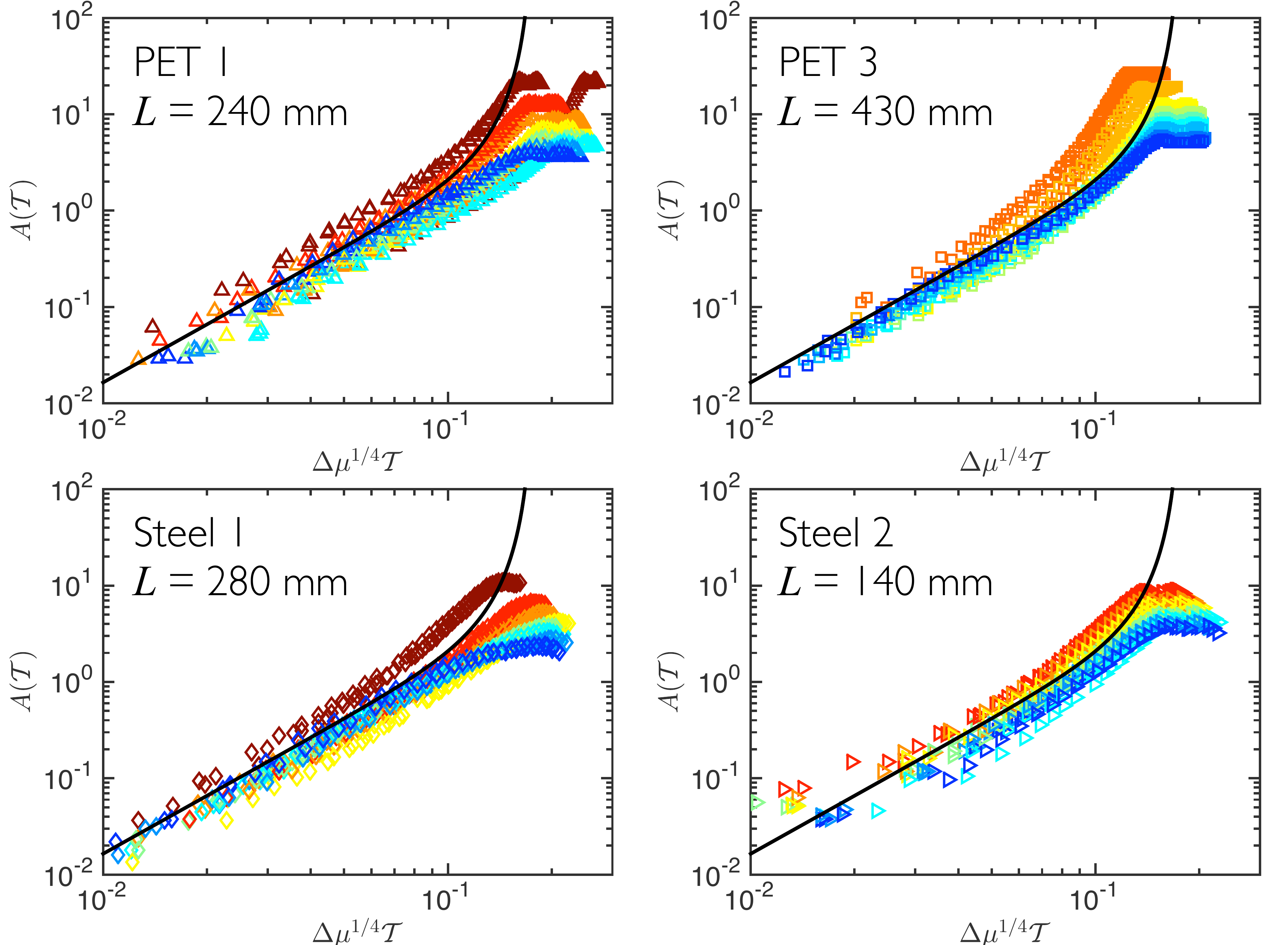}  
\caption{Dimensionless midpoint trajectories during snapping for PET and steel strips of varying natural length $L$. In each plot, different trajectories correspond to different end-shortenings beyond the snapping transition, coloured as a heat-map from small values of $\Delta\mu$ (brown/red) up to large values of $\Delta\mu$ (blue). Data is shown for PET strips  with $L = 240~\mathrm{mm}$, $\alpha = 21.34^{\circ} $ (upward-pointing triangles) and $L = 430~\mathrm{mm}$, $\alpha = 21.17^{\circ} $ (squares), as well as for experiments on steel strips with $280~\mathrm{mm}$, $\alpha = 17.14^{\circ} $ (diamonds) and $L = 140~\mathrm{mm}$, $\alpha = 22.51^{\circ}$ (right-pointing triangles). }
\label{fig:othercollapse}
\end{figure}

The results  in figure \ref{fig:othercollapse} show that there is a reasonable collapse of experimental data with different values of $\dL$ onto a single curve for $A(\tnd)$. However, we note that there is some dispersion of the data about the numerically predicted $A(\tnd)$. We attribute this dispersion to uncertainty in the measured value of $\dL$ ($\pm 200~\mu\mathrm{m}$), which increases the relative error in $\Dmu$ as $\dL\to\Lf$.


\end{document}